# Predissociation dynamics of negative ion resonances of $H_2$ near 12 eV and 14.5 eV using velocity slice imaging technique

Akshay Kumar[1], Suvasis Swain[1], Jankee Upadhyay[2], Yogesh Upalekar[1], Rajesh Arya[2], and Vaibhav S. Prabhudesai[1]*

[1]*Tata Institute of Fundamental Research, Colaba Mumbai 400005 India*

[2]*Laser Electronics Division, Raja Ramanna Centre for Advanced Technology, Indore 452013 India*

*vaibhav@tifr.res.in

**Abstract**

Dissociative electron attachment (DEA) is an important tool for investigating negative ion resonances. We have studied the negative ion resonances of $H_2$ at 10 eV and 14 eV using the improved velocity slice imaging technique. We obtained modulations in the kinetic energy spectrum of $H^-$ ions obtained at 12 eV and 14.5 eV electron energy, consistent with the earlier reported vibrational state contributions from the higher lying bound resonances. We show that structures obtained at 12 eV are due to predissociation of the $C\,^2\Sigma_g^+$ resonance consistent with the current understanding. However, based on our angular distribution measurements, we propose that the structures obtained at 14.4 eV are due to predissociation of bound resonance of $^2\Sigma_g^+$ symmetry as against $\Delta_g$ that was proposed earlier. We also report that the bound $^2\Sigma_g^+$ resonance contributes to the observed inversion symmetry breaking near 14 eV.

1. Introduction

Dissociative electron attachment (DEA) is a useful probe to study negative ion states. In DEA, the electron is resonantly attached to a molecule to form temporary negative ion resonances (NIRs), which can dissociate into an anion and one or more neutral fragments. Although electron collisions have been studied for almost a century, many new facets of molecular NIRs are unraveling due to the invention of advanced experimental methods like momentum imaging [1]. For example, only recently, Krishnakumar *et al*. have provided reliable values of the experimentally measured absolute cross-section of DEA to the simplest molecules, $H_2$ and $D_2$, using the momentum imaging technique [2]. Here, velocity slice imaging (VSI), one of such techniques adopted for these measurements, helped to eliminate the contribution from the electronically excited metastable neutrals and ultraviolet light while ensuring the detection of all the ions. These cross-sections have provided a benchmark to test the theoretical tools invented [3,4]. From a theoretical aspect, barring a few exceptions, molecular NIRs still need to be successfully modeled. Even for $H_2$, NIRs are not entirely understood. The observation of quantum coherence in the DEA process in $H_2$ and $D_2$ at 14 eV is another example of the new facets of NIRs unraveled by the VSI technique [5]. These results highlighted the role of the $^2\Sigma_u^+$ state at these energies, which has not been identified by any theoretical calculations. On the other hand, Swain *et al.* have shown that from 10.5 eV onwards, the $B^2\Sigma_g^+$ negative ion resonance state of $H_2$ lies below the parent $b^3\Sigma_u^+$ state [6], which is in contrast with the latest theoretical calculations [7].

$H_2$ is the simplest and ideal molecular system to benchmark the theoretical models that describe excited NIRs and their dissociation. Many studies have been carried out on this system in the last few decades [8-10]. DEA to $H_2$ is of fundamental importance. It leads to the formation of hydride ion that plays a significant role in the chemistry of the interstellar medium [11-13] and fusion plasmas [14-16]. The $H^-$ signal from $H_2$ in a low energy regime (<17 eV) comes from three resonant processes [2, 17]. The first resonant process occurs around 4 eV via the lowest attractive ground state $X^2\Sigma_u^+$ dissociating into $H(^2S)$ and $H^-(^1S)$. This is a threshold process that occurs from the dissociation limit of the ground anion state (3.75 eV). The second resonant process occurs in the 8 to 13 eV energy range and is associated with the repulsive $B^2\Sigma_g^+$ resonant state and leads to $H(^2S)$ and $H^-(^1S)$ products with high kinetic energy. The third resonant process around 14 eV, which was earlier believed to be associated with only $^2\Sigma_g^+$ resonance [18], is found to be associated with the coherent superposition of two resonant states of $^2\Sigma_g^+$ and $^2\Sigma_u^+$ symmetry, which causes forward-backward asymmetry in the angular distribution [5]. The 14 eV peak leads to $H^-(1s^2)$ and $H(n = 2)$ dissociation products. As the dissociation limit for this process is 13.95 eV, the $H^-$ is formed with low kinetic energies, similar to that of the 4 eV channel.

The second $B^2\Sigma_g^+$ resonance was studied earlier. In the high electron energy resolution experiment, Dowel and Sharp [19] observed structures in $H^-$ signal intensity as a function of electron energy in the 11.3 eV-13.3 eV range. Tronc *et al.* [20] have also reproduced these structures. These structures, periodic in electron energy, were in good agreement with the positions of the vibrational levels of resonance series "A" (as per the notation of Schulz *et al.* [21]) observed in the electron ejection channel [22-24]. This agreement suggests that structures are due to $C^2\Sigma_g^+(1s\sigma_g)^1(2p\sigma_u)^2$ resonant state. The structures can be understood due to the predissociation of the vibrational levels of attractive $C^2\Sigma_g^+$ via the repulsive $B^2\Sigma_g^+$ state. Swain *et al.* [6] have observed the effect of $C^2\Sigma_g^+$ resonance in the angular distribution of $H^-$ from $H_2$ in the 8-13 eV region. However, due to the poor energy resolution of the system, they were unable to get direct evidence of these structures in the ion yield spectrum.

For the 14 eV peak in the DEA yield, it has been shown that the electron attachment to the ground state ($X^1\Sigma_g^+$) of $H_2$ leads to the formation of a coherent superposition of $^2\Sigma_g^+$ and a $^2\Sigma_u^+$ resonant states [5]. It involves the simultaneous transfer of $s$ and $p$ partial waves from the attaching free electron breaking the inversion symmetry in the system. This symmetry breaking leads to forward-backward asymmetry in angular distribution. However, in the earlier measurements by Tronc *et al.* [25] with higher electron energy resolution, modulations similar to the 10 eV peak were observed in the H⁻ ion-yield curve as a function of electron energy [25]. In electron transmission experiments, structures at these energies were identified as "*band f*" and were observed by Weingartshofer [26], Golden [27], and Sanchez and Schulz [24]. These authors assigned the $^2\Sigma_g^+$ symmetry for the '*f*' band. However, Tronc *et al.* [25] have assigned $\Delta_g$ symmetry for corresponding resonance and proposed that these structures are due to the interaction of $\Sigma_g$ and $\Delta_g$ states through rotations. In this context, it is worthwhile to investigate the features observed at 14 eV using the improved VSI technique, which can show the contribution from the vibrational states of the bound resonance.

The vibrational structures observed in both 10 eV and 14 eV DEA peaks were obtained using a high-resolution electron beam in the DEA cross-section measurements [20, 25]. So far, they have not yet been reported in the momentum imaging measurements due to the limited electron energy and ion momentum imaging resolution of the apparatus used. In this work, we report these vibrational structures with the improved resolution of momentum imaging set up for both 10 eV and 14 eV peaks in DEA measurements. Furthermore, using this improved spectrometer, we have obtained details of the symmetry of involved resonances.

## 2. Experimental setup

A magnetically collimated electron beam generated by thermionic emission from the heated tungsten filament was crossed with the effusive molecular beam produced using a capillary array. The electron gun was operated in the pulsed mode (width 100 ns and repetition rate 3000 Hz). The electrons were collected by the Faraday cup situated co-axially at the other end of the interaction zone of the VSI spectrometer. The electron beam was collimated using the magnetic field of 50 Gauss generated using a pair of coils mounted in the Helmholtz geometry outside the vacuum chamber. The interaction volume spanned by the overlap of the electron, and the molecular beam was situated at the center of the interaction region of the VSI spectrometer. The interaction region of the spectrometer was flanked by the pusher and puller electrodes separated by 20 mm. The puller electrode has a central aperture of 30 mm lined with a molybdenum wire mesh of 64% transmission. The negative ions formed from electron molecule interaction were extracted into the lens region using a delayed pulsed extraction field. A square voltage pulse of -60 V amplitude and 1 μs duration was used as the extraction pulse on the pusher electrode, which was delayed by 90 ns with respect to the electron pulse. The energy resolution of the electron gun was about 1 eV. The chamber was pumped by oil-free pumps to a base vacuum of $1 \times 10^{-8}$ Torr. During the experiments, the base pressure of $H_2$ gas was kept at $3 \times 10^{-6}$ Torr, and the electron current was 0.36 nA. The energy calibration of the electron beam was carried out by observing the 14 eV peak of H⁻ from $H_2$.

The generated ions were velocity focused via a 4-lens assembly [28]. This 4-lens assembly allowed the momentum images to be zoomed for low-energy ions to improve the imaging resolution. The ions were detected by a 2D position-sensitive detector (PSD) mounted at the end of the flight tube. The detector was made of two 75 mm diameter active area microchannel plates (MCP) mounted in the Chevron configuration, followed by a phosphor screen. The images formed on the phosphor screen were recorded by a charge-coupled device (CCD) camera. The VSIs obtained were then analyzed after adding several such slices in the offline analysis.

The detector was kept active only when the central slice of the Newton sphere arrived at it. The appropriate delay of the central slice was obtained by shifting the detector activation window with respect to the pusher pulse and by obtaining the image with the maximum radius. The biasing of the detector was obtained using the Behlke switch having a pulse duration of 10 ns synchronized with the electron pulse with a suitable delay with respect to the pusher pulse.

The Behlke switch-based pulse generator is specifically developed for the experimental setup. The requirement was to generate a high voltage (HV) pulse of 2.5 kV with a pulse duration of 10 ns and a switching time of better than 2 ns at an adjustable repetition rate of more than 1 kHz. There are several techniques for achieving HV fast switching [29-31] based on switches realized by using Transistors, MOSFETs, Insulated Gate Bipolar Transistors (IGBTs), etc. In the present work, the pulse generator has been developed using a Behlke solid-state HV switch (HTS-50-08-UF). These solid-state switches are specially designed and developed to generate HV pulses of a short duration of 10 ns, and a constant fast rise time better than 2 ns. These switches, being semiconductor devices, have a very low turn-on jitter of the order of 100 ps and a longer lifetime. These switches can provide galvanic isolation of more than 10 kV, thus can be floated at the required high potential and used as high-side switches for positive as well as for negative voltages.

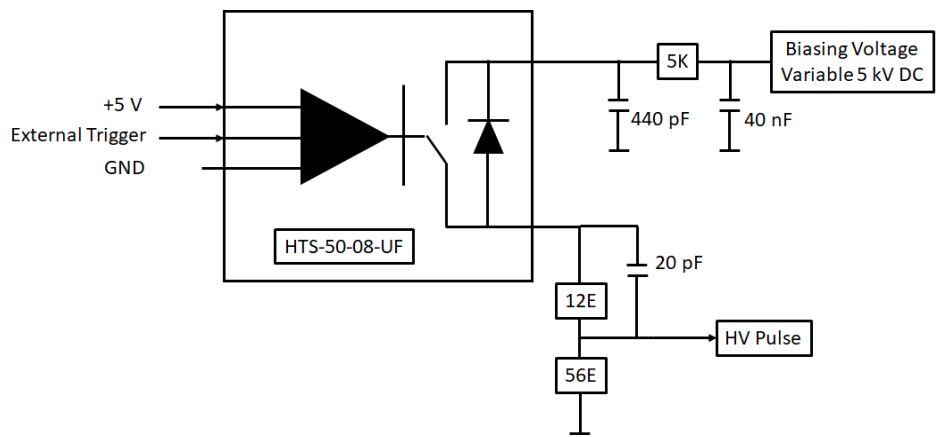

Fig.1: Block diagram of the high-voltage pulse generator unit

The block diagram of the pulse generator is shown in Fig. 1. It has been realized by charging a 440 pF/6 kV capacitor to the desired voltage and discharging into a load resistance of matched impedance using HV switching in synchronization with an external trigger signal. The discharging current generates an HV pulse of half the applied biasing voltage with a duration of 10 ns fixed for the selected Behlke switch. The pulse amplitude can be varied by varying the applied biasing voltage. High-speed switching often leads to various difficulties including self-oscillating, self-re-triggering, ringing, etc. In order to overcome these difficulties, a printed circuit board (PCB) layout has been designed by minimizing the value of parasitic reactive components, matching the load with output impedance, and implementing the star grounding scheme. EMI shielding of the pulser circuit has been achieved by specifically designed and fabricated enclosure, of size 220 mm X 110 mm X 185 mm using 3 mm MS Ni-Chromium Plated Sheet.

For the 10 eV resonance, the threshold for H⁻ formation is 3.75 eV. This implies that at 12 eV electron energy, the KE of fragments would be about 4 eV. Due to low mass and high kinetic energy, momentum imaging H⁻ from $H_2$ at 10 eV is particularly challenging. The magnetic field applied for electron collimation also adversely affects the H⁻ imaging. Due to Lorentz force, one side of the newton sphere moves close to the edge of the lens electrode's aperture, which results in distortion of momentum images. However, due to cylindrical symmetry around the electron beam, we can use only the other half of the image that passes close to the spectrometer axis to obtain the characteristic dissociation dynamics. In addition, due to the presence of static gas background, it produces an extended source of ions that causes distortions in images [4]. To remove these distorting features, static gas background subtraction was done by diverting the gas flow through another entrance and keeping the base pressure the same.

The energy resolution of the momentum image is directly related to the annular width of the obtained image. The annular width depends upon the velocity focusing conditions, time slicing width, and the electron's energy resolution width. To determine the focussing conditions, the potentials on various electrodes were estimated using SIMION simulations. The electrode voltages were further fine-tuned in the experiment to obtain the best achievable velocity-focusing condition. The small slicing width can reduce the contribution of the non-central part of the Newton sphere and thus can help in improving the energy resolution of the image. Poor electron energy resolution has an advantage for $H_2$. As a diatomic molecule, on dissociation from a particular resonance, excess energy above the threshold of the process would appear as the fragment's kinetic energy (KE). We can access the different parts of the ion yield curve over the energy spread in the electron beam by fixing the mean electron's energy at a particular value. We used SIMION to obtain the expected energy resolution of the momentum image for a specific set of the initial kinetic energies and with the other conditions matching the experimental scheme. The effect of the thermal motion of particles from an effusive jet having an aspect ratio of 10 at room temperature, 1 cm from the end of the capillary array, and dissociating in random directions in the interaction region, was estimated. It was found that the thermal motion was adding to the energy width by about 25 meV. This spread of energy was incorporated into the SIMION simulations. The ions were flown, and the Newton sphere was time sliced in a separate analysis. The annular width of the obtained image is used to determine the corresponding momentum resolution of the imaging condition. Due to the presence of the magnetic field, the image loses the cylindrical symmetry about the spectrometer axis. As a result, energy resolution was found to be angle-dependent. This effect is more pronounced for high KE ions. For 4 eV, H⁻ ions energy resolution is found to be 80 meV near 30°-60° and around 90 meV around 40°-70° with respect to incident electron direction for 10ns time gating and about 300 meV for 80 ns time gating.

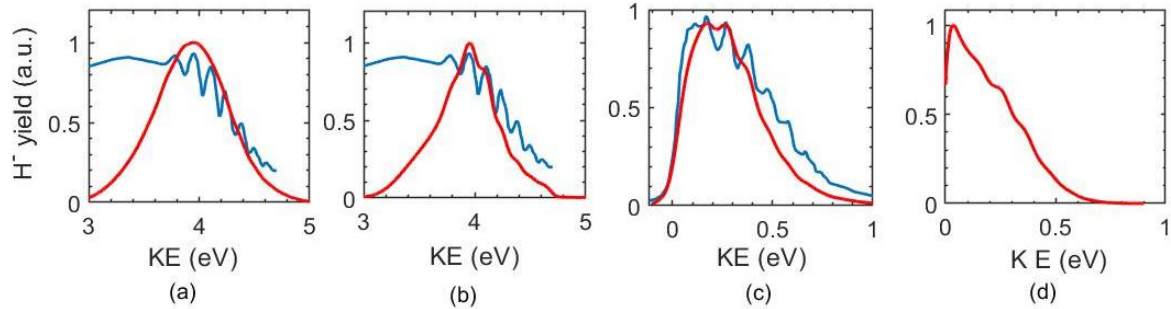

Fig.2: Expected KE distribution (solid red line) H⁻ from $H_2$ around 10 eV for different time slicing conditions obtained after multiplying the ion yield data (solid blue line) from ref (20) with the electron gun profile centered around 12 eV and convoluting the product with (a) 300 meV imaging resolution obtained for 80 ns time gating and (b) 90 meV imaging resolution obtained for 10 ns time gating. Expected H⁻ KE distribution (solid red line) from $H_2$ around 14 eV electron energy range obtained after multiplying the ion yield data (solid blue line) from ref (25) with the electron gun profile of 1 eV FWHM centered around 14.4 eV and convoluting the product with (c) 60 meV imaging resolution, and (d) after taking into account weight factor (see the text).

To compare our data with that from Tronc *et al*. [20], we multiplied their reported cross-section data with the electron gun profile of our experiment and convoluted the result with an obtained energy resolution of the imaging spectrometer. Fig. 2 shows the importance of 10 ns time gating compared to 80 ns time gating. Despite poor electron energy resolution, we could observe the effect of the vibrational structures in KE distribution in the 10 ns slicing due to good imaging resolution.

The VSI condition was modified to map the low-energy H⁻ ions obtained at 14.4 eV with appropriate magnification. Under that lensing condition, SIMION simulations were carried out to determine the imaging resolution of the spectrometer. After a similar analysis as done for 4 eV H⁻ ions, we found energy resolution for 0.3 eV H⁻ ions to be around 60 meV in all angular ranges. To obtain the expected KE distribution, we multiplied the observed ion yield data of Tronc *et al*. [25] with the electron gun profile and convoluted the result with the 60 meV energy resolution of the imaging spectrometer. The obtained spectrum is shown in Fig. 2(c). For a blob, counts near 0 eV are artificially enhanced due to the geometry factor arising from the finite width of the slice, which gives an intrinsic bias towards the low-energy ions. The simulated KE distribution obtained after multiplying the data from Tronc *et al*. near 14 eV [25] with the electron gun profile needed to be multiplied by an additional weight factor that depends on the overall spread of the time of flight peak against the width of the slicing. After multiplying the weight factor and convoluting it with 60 meV imaging resolution, the modified KE distribution is shown in Fig. 2(d).

## 3. Results and Discussion

For a homonuclear diatomic molecule, due to the presence of inversion symmetry in the system, electron capture proceeds through the transfer of either odd or even partial waves [32]. Under the axial recoil approximation, the angular preference during electron capture will be mapped in the angular distribution of dissociation products. The general form of the angular distribution [33] of the molecule is given by

$$I(k,\theta,\varphi) = \left|\sum_{l=m}^{\infty} A_{lm}(k)Y_{lm}(\theta,\varphi)\right|^2 \qquad (1)$$

where $A_{lm}$ is the transition amplitude between the target state and the resonant state, $l$ is the orbital angular momentum of the incident electron, and $m$ is the difference between the electronic axial angular momenta of the target state and resonant state. $Y_{lm}(\theta,\varphi)$ are the corresponding spherical harmonics involved, angle $\theta$ represents the angle of ejection of an anion fragment with respect to the incident electron beam, and $\varphi$ is the corresponding azimuthal angle.

### 3.1 Predissociation near 12 eV

The experimentally obtained momentum image for the H⁻ ions from H$_2$ at 12 eV after background subtraction is shown in Fig. 3(a). At 12 eV, the blob seen in the center of the image is due to the long energy tail of the electron beam, which produces negative ions from the 14 eV resonance. The observed shift of the blob away from the center is due to the effect of the magnetic field used for electron beam collimation [6]. This is consistent with the simulations carried out with the charged particle trajectory simulation program: SIMION. The thin ring obtained is from the 4 eV ions formed from the DEA via $B^2\Sigma_g^+$ resonance. The effect of thin slicing of 10 ns can be seen in the annular width of the ring as compared to earlier 80 ns slicing [6]. Near 12 eV, the transition occurs between the $X^1\Sigma_g^+$ state of H$_2$ to the repulsive $B^2\Sigma_g^+$ and bound $C^2\Sigma_g^+$ states of H$_2^-$ [19]. According to the selection rule of $g \to g$ transition, only even partial waves are allowed. At low energy, lower allowed partial waves have a dominant contribution to the capture.

Therefore, at 12 eV, the contribution from only $s$ and $d$ partial waves of the electron would suffice to describe the angular distribution. A similar analysis will also hold for the upper bound With the $C^2\Sigma_g^+$ resonance. The angular distribution from the corresponding transition can be expressed as

$$I(k,\theta) = \left|A_{00}(k)Y_{00}(\theta,\varphi) + A_{20}(k)Y_{20}(\theta,\varphi)e^{-i\delta}\right|^2 \qquad (2)$$

which on solving gives

$$I(k,\theta) = \frac{A_{00}^2(k)}{4\pi} + 5\frac{A_{20}^2(k)}{16\pi}(3\cos^2\theta - 1)^2 + \sqrt{5}\frac{A_{00}(k)A_{20}(k)}{4\pi}(3\cos^2\theta - 1)\cos(\delta) \qquad (3)$$

Here $A_{00}$ and $A_{20}$ are transition amplitudes corresponding to the attachment of $s$ and $d$ waves of incident electron, respectively, and $\delta$ is the relative phase between two partial waves.

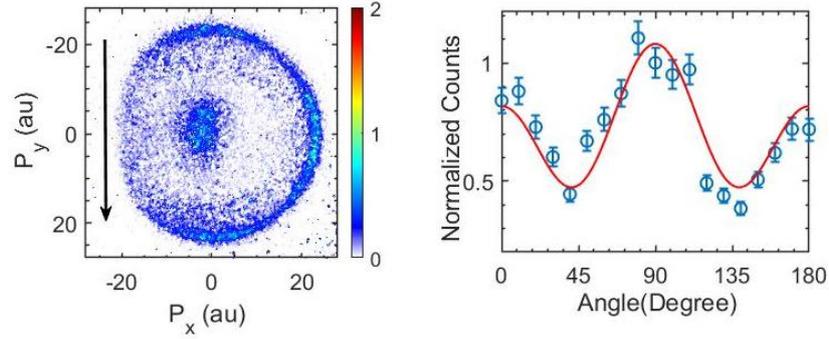

Fig. 3: (a) Background subtracted momentum image of H⁻ from $H_2$ obtained at 12 eV electron energy. The direction of the electron beam is from top to bottom, as indicated by the black arrow. (b) Angular distribution of the ions in the KE range of 3-5 eV and the corresponding fitting of angular distribution is also shown (solid red line). The angular distribution is normalized at $90°$.

Using equation 3, the angular distribution obtained for the momentum image (Fig. 3(a)) is fitted. The corresponding fitted curve is shown by the solid red line in Fig. 3(b). Using the fitted parameters ratio of transition amplitudes ($A_{00}/A_{20}$) is calculated, and the ratio is found to be $1.9 \pm 0.1$, which is consistent with earlier reported results [6]. The increased $s/d$ ratio around 12 eV is explained due to additional contribution from the resonance of identical symmetry ($C^2\Sigma_g^+$) [6].

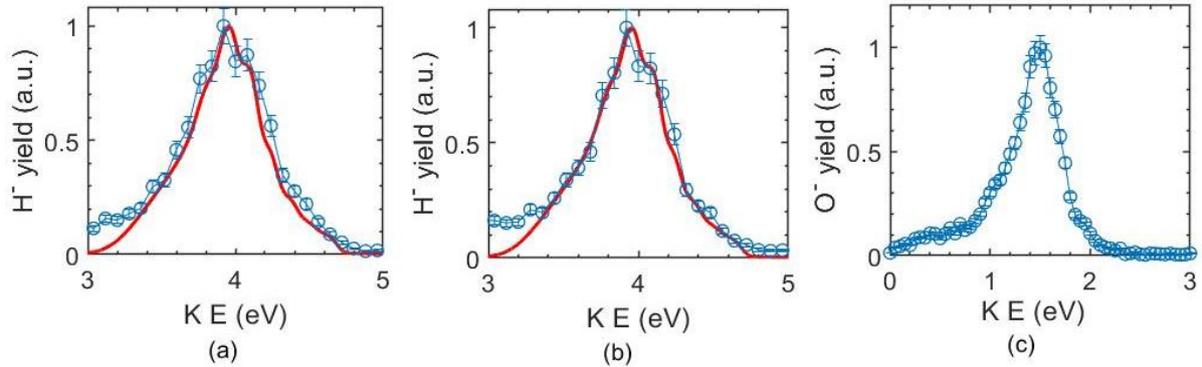

Fig. 4: KE distribution of H⁻ from $H_2$ obtained at 12 eV electron energy in the angular range of (a) $40° - 70°$ (b) $30° - 60°$ using time slicing of 10 ns. The effect of thin-slicing can be seen in vibrational structures in the ion yield spectrum. The ion yield curve obtained after multiplying the data from Tronc *et al*. [20] with the electron gun profile and convoluting it with imaging resolution (see the text) is also shown in a solid red line. (c) KE distribution of O⁻ from $O_2$ obtained around 6.5 eV electron energy around $90°$ using time slicing of 10 ns and the same spectrometer operating condition.

To see the effects of $C^2\Sigma_g^+$ resonance on the ion yield spectrum, KE distribution of H⁻ ions was plotted in the region where the contribution due to the $d$ wave is minimum, i.e., around $55°$. The obtained KE distribution for the 12 eV images in the $40° - 70°$ angular region is shown in Fig. 4(a). The corresponding expected KE distribution from Fig. 2(b) is also shown as the solid red line. The presence of vibrational structure can be seen as shoulders at 3.8 eV and 4.2 eV, which correspond to peaks coming at 11.3 eV and 11.9 eV, respectively, in the ion yield curve reported by Tronc *et al*. [20]. Similar effects can be seen for KE distribution plotted in a $30° − 60°$ angular region for which the best energy resolution is expected (80 meV), as shown in Fig. 4(b). The counts before 3.5 eV were generated due to the remaining background contribution. To verify that these structures are not from any experimental artifact, we have also obtained the VSI image of $O^-$ from $O_2$ using 10 ns slicing. For $O_2$, we don't expect any structures in the kinetic energy spectrum. As shown in Fig. 4(c), the obtained KE spectrum of $O^-$ from $O_2$ is free from any structures and thus verifies that the structures observed in $H_2$ are not imaging artifacts.

### 3.2 Predissociation near 14 eV

After obtaining the earlier reported vibrational structure in the KE distribution at 12 eV, we carried out similar measurements for the 14 eV peak. Since the KE of $H^-$ at 14.4 eV is < 0.7 eV, the corresponding VSI image is a blob, as shown in Fig. 5(a). The energy calibration is carried out from the 15 eV image. The corresponding KE distribution is shown in Fig. 5(b), along with the expected curve from Fig. 2(d). Earlier, 14 eV resonance was believed to be associated with only $^2\Sigma_g^+$, but recently it was shown that 14 eV resonance is associated with both $^2\Sigma_g^+$ and $^2\Sigma_u^+$ resonance. Thus, structures observed around 14 eV in principle can come from the interaction of both resonances with another high-lying bound resonance. To check the involvement of the resonances, the KE distribution of $H^-$ for the 14.4 eV electron energy image was obtained around the $90°$ angular region with respect to the incoming electron beam direction and is plotted in Fig. 5(c).

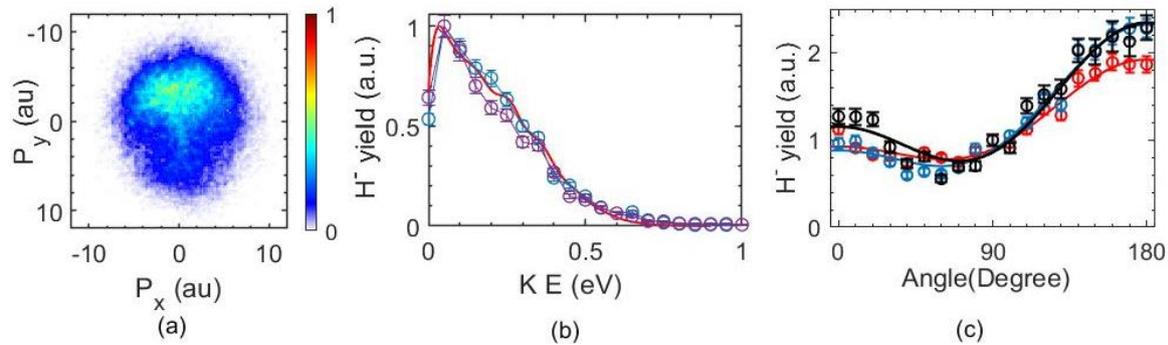

Fig. 5: (a) Background subtracted momentum image of $H^-$ from $H_2$ obtained at 14.4 eV electron energy using the 10 ns time slice. The direction of the electron beam is from top to bottom (b) corresponding KE distribution obtained in the angular range of $75° − 105°$ (blue circles) and $40° − 70°$ (violet circles). The solid red line shows the expected KE distribution shown in Fig 1(d). (c) The comparison of the angular distribution obtained from the image and normalized at $90°$ in the KE range of 0-0.15 eV (red circles), where we expect no vibrational structures with the KE ranges of 0.15-0.3 eV (blue circles) and 0.3-0.45 eV (black circles) where we expect vibrational structures. The solid lines show the corresponding fits.

Since the contribution of $^2\Sigma_u^+$ is minimum around $90°$, the presence of vibrational structures around $90°$ shows that they are arising due to the interaction of $^2\Sigma_g^+$ with another bound resonance. Tronc *et al.* [25]

have also observed the structures around 90°, and it was proposed that these structures are due to the interaction of $^2\Sigma_g^+$ with $^2\Delta_g$ state through rotation coupling.

To determine the symmetry of bound resonance, we have plotted the angular distribution in different KE ranges. KE distribution has no structures for 0-0.15 eV, as shown in Fig. 5(b). The angular distribution in this region would be due to the capture of only $s$ and $p$ partial waves of a free electron coming from transition $X^1\Sigma_g^+ \to {}^2\Sigma_g^+$ and $X^1\Sigma_g^+ \to {}^2\Sigma_u^+$ respectively. The expected angular distribution can be expressed [34] as

$$I(k,\theta) = \left|A_{00}(k)Y_{00}(\theta,\varphi) + A_{10}(k)Y_{10}(\theta,\varphi)e^{-i\delta}\right|^2 \qquad (4)$$

$$I(k,\theta) = A_{00}^2(k) + 3A_{10}^2(k)\cos^2\theta + 2\sqrt{3}A_{00}(k)A_{10}(k)\cos(\theta)\cos(\delta) \qquad (5)$$

where $A_{00}$ and $A_{10}$ are transition amplitudes corresponding to the attachment of $s$ wave and $p$ wave of incident electron, respectively, $\delta$ is the sum of the relative phase gained during the dissociation along the two paths, and the initial relative phase between the $s$ and $p$-waves. The presence of the $cos(\theta)$ term will induce forward-backward asymmetry in the angular distribution.

Electron capture from $X^1\Sigma_g^+$ state of H$_2$ to $\Delta_g$ resonant state requires the capture of $d$ wave of a free electron. However, the angular distribution shown in Fig. 5(c) obtained for the 0.15-0.3 eV and 0.3-0.45 eV is similar to the 0-0.15 eV KE range and could be fitted to equation (5). Obtained fitted parameters for angular distribution in different KE ranges are shown in Table 1. This indicates that there is no involvement of $d$ wave in the angular distribution; hence, electron capture would not be happening to a $\Delta_g$ state.

**Table 1.** Transition amplitudes and relative phase from the angular distributions H⁻ from H$_2$ at 14.5 eV electron energy in different KE ranges.

| KE range (eV) | $A_{00}$ | $A_{10}$ | $\delta(rad)$ |
|---|---|---|---|
| 0 – 0.15 | 0.94 ± 0.07 | 0.42 ± 0.08 | 4.34 ± 0.10 |
| 0.15 – 0.3 | 0.94 ± 0.04 | 0.49 ± 0.05 | 4.24 ± 0.06 |
| 0.3 – 0.45 | 0.93 ± 0.08 | 0.54 ± 0.06 | 4.36 ± 0.06 |

Moreover, the Contribution of $d$ partial wave is minimum around 55° if the bound resonance is of $\Delta_g$ symmetry. Hence, we should not expect any structures in KE distribution around 55°. On the contrary, we have observed structures in KE distribution in the form of shoulders appearing at the 0.15-0.3 eV KE range and 0.3-0.45 eV KE range. around 55° similar to structures that were observed around 90°, as shown in Fig. 5(b). The corresponding KE distribution is also compared with the simulated distribution (Fig. 1(d)) and is shown in the solid red line in Fig. 5(b). This eliminates the possibility of the higher lying bound resonance being of the $\Delta_g$ symmetry. Sanchez and Schulz have also observed 14 eV structures in the electron scattering experiment and assigned $\Sigma_g$ symmetry to the corresponding resonance. From these observations, we conclude that the vibrational structures observed in the ion yield of the 14 eV resonance are arising due to the interaction of $^2\Sigma_g^+$ resonance with another high-lying bound resonance of $^2\Sigma_g^+$ symmetry. Using the energy values where Tronc *et al*. [25] have got the structures in the 14 eV region and comparing it with vibrational states [35] of H$_2$, we proposed that high-lying bound $^2\Sigma_g^+$ resonance originates from $D'^1\Pi_u$ state with dissociation limit H + H(n = 4) of H$_2$ as the parent state [36].

To observe the effect of bound $^2\Sigma_g^+$ resonance that is causing structures in the 14 eV region on the symmetry breaking, we have determined the asymmetry parameter [34] of velocity slice image obtained at 14.4 eV in different KE ranges, as shown in Table 2. From earlier calculations [5], it was shown that if we consider only $^2\Sigma_g^+$ and $^2\Sigma_u^+$ dissociating resonances, the energy-integrated asymmetry parameter decreases with electron energy beyond 14 eV. On the other hand, if bound $^2\Sigma_g^+$ resonance contributes incoherently to the DEA process, the asymmetry parameter should decrease in the 0.15-0.45 eV KE range where there is a contribution from the bound $^2\Sigma_g^+$. But from Table 2, it can be seen that the involvement of bound $^2\Sigma_g^+$ is causing an increase in asymmetry. This shows that high lying $^2\Sigma_g^+$ resonance is coherently participating in the symmetry-breaking process observed near 14 eV. It is also important to note that the next dissociation limit available for the DEA process, namely H(n=3) + H$^-$($^1$S), is at 15.86 eV [37], which is just about at the edge of the energy spread of the electron beam at 14.4 eV. However, the measured cross-section at this energy is very small [2]. Consequently, this channel will have a negligible contribution to the signal in the KE range of 0-0.15 eV and will not influence the inference.

**Table 2:** Comparison of measured asymmetry parameter ($\eta$) in different KE ranges of H$^-$ from H$_2$ at 14.4 eV electron energy.

| KE range (eV) | Experimental values of $\eta$ |
|---|---|
| 0 – 0.15 | -0.31 ± 0.02 |
| 0.15 – 0.3 | -0.38 ± 0.02 |
| 0.3 – 0.45 | -0.35 ± 0.02 |

4. **Conclusion**

Using a VSI spectrometer with improved energy resolution due to 10ns time slicing, we have observed the effect of vibrational structures at 12 eV and 14.4 eV electron energy in the kinetic energy distribution of H$^-$ ions from H$_2$. These structures are consistent with the earlier reports by Tronc *et al*. [20, 25]. At 12 eV, the structures are due to predissociation of bound $C^2\Sigma_g^+$ resonance state via the repulsive $B^2\Sigma_g^+$ resonance. The main contribution to the repulsive resonance is from the transfer of $s$ and $d$ partial waves to the target molecule. The bound resonance contributes predominantly to the $s$ wave capture. At 14.4 eV, we identify the structures in the KE distribution of ions as due to the predissociation of $^2\Sigma_g^+$ bound resonance in contrast to the earlier proposed $^2\Delta_g$ resonance state. We infer it from the angular distributions obtained at various KE ranges and by obtaining the KE distributions in various angular ranges, ruling out the contribution of the $d$ partial wave. We propose that this bound resonance of $^2\Sigma_g^+$ symmetry may have the $D'\,^1\Pi_u$ state of neutral H$_2$ as the parent state. We also propose that this upper bound resonance contributes to the symmetry breaking of the inversion symmetry as its contribution should be added coherently to the resultant transition.

**Acknowledgement**

SS, AK, and VSP acknowledge the financial support from the Department of Atomic Energy, India, under Project Identification No. RTI4002. All authors acknowledge Dr. S. V. Nakhe for support and Mr. Sudhir Kumar for help in fabrication of the pulse generator.